\documentclass[11pt]{article}
\usepackage[margin=1in]{geometry}
\usepackage{amsmath,amssymb}
\usepackage{graphicx}
\usepackage{booktabs}
\usepackage[colorlinks=true,citecolor=blue,linkcolor=blue,urlcolor=blue]{hyperref}
\usepackage{authblk}

\title{\bf From Exact Diagonalization to DMRG: A Complete Numerical Study of the Transverse-Field Ising Model}
\author{Chandra Sekhar Prayaga}
\affil{Department of Physics, University of West Florida, Pensacola, FL 32514}
\date{}

\begin{document}
\maketitle

\begin{abstract}
\noindent
We present a self-contained numerical study of the one-dimensional transverse-field Ising model (TFIM), tracing its ground-state entanglement structure from exact diagonalization at small system size ($L=8,14,20$) through density-matrix renormalization group (DMRG) calculations up to $L=100$. A single, consistent methodology --- identical field grid, identical convergence diagnostic (independent runs at bond dimensions $\chi=100$ and $\chi=200$ at every point) --- is used across all seven system sizes, and we show explicitly where exact diagonalization and DMRG must agree exactly rather than merely approximately. Along the way we document and resolve a bond-dimension convergence artifact that produced a spurious discontinuity in an earlier, less systematic dataset, as a worked cautionary example for practitioners. Using the resolved dataset we extract the central charge of the transition via two complementary routes --- the leading logarithmic finite-size scaling of the mid-chain entropy, and the full Calabrese-Cardy formula applied to every bond of every system size simultaneously --- obtaining $c_{\rm eff}\to 0.51$--$0.52$ as short-distance lattice corrections are systematically excluded, consistent with the exact two-dimensional Ising value $c=1/2$. We review the quantum-classical (Suzuki-Trotter) correspondence that underlies this agreement and discuss how the Widom-Kadanoff scaling hypothesis transplants from the classical to the quantum problem. The manuscript is intended as both a physics result and a worked methodological example of careful finite-size numerics.
\end{abstract}

\section{Introduction}

The transverse-field Ising model (TFIM) is among the simplest quantum many-body Hamiltonians that exhibits a continuous quantum phase transition, and for exactly that reason it has served for decades as the standard testbed for new numerical and analytical techniques in quantum many-body physics. In one spatial dimension it is exactly solvable via a Jordan-Wigner mapping to free fermions \cite{Pfeuty1970}, its transfer matrix is identical to that of the two-dimensional classical Ising model in an appropriate limit \cite{Fradkin1978}, and its critical point lies in a universality class --- the two-dimensional Ising / free Majorana fermion CFT, with central charge $c=1/2$ --- that is by now textbook material. None of this makes the model uninteresting to study numerically: precisely because the answer is known, the TFIM is an ideal proving ground for the numerical techniques (exact diagonalization, DMRG) that will later be applied to less tractable models, and for the finite-size and convergence subtleties that such techniques inevitably encounter.

This note has two aims, reflected in its structure. First, it presents a complete, internally consistent numerical dataset for the TFIM ground-state entanglement entropy across seven system sizes, $L=8$ through $L=100$, produced with a single method throughout so that no system size needs to be treated as a special case. Second, and in the spirit of a Lecture Notes format, it documents the numerical practice as carefully as the physics: where exact diagonalization and DMRG are guaranteed to agree exactly rather than approximately (Section~\ref{sec:exact}), how a genuine convergence artifact was identified and resolved in an earlier iteration of this dataset (Section~\ref{sec:artifact}), and how a central-charge extraction can be strengthened by using the full bond-resolved entropy profile rather than a single system-size scan (Section~\ref{sec:cc}).

We proceed bottom-up: Section~\ref{sec:model} reviews the model, the entanglement-entropy diagnostic, and the quantum-classical correspondence that explains why the 1D quantum problem inherits 2D classical Ising critical behavior. Section~\ref{sec:methods} describes the two numerical methods and the point at which the Hilbert space forces the transition from one to the other. Section~\ref{sec:results} presents the results: finite-size flow of the entropy, the bond-resolved entropy profile, and the central-charge extraction. Section~\ref{sec:discussion} discusses the results and situates them relative to known subtleties in the literature; Section~\ref{sec:conclusion} concludes.

\section{Model and Theoretical Background}
\label{sec:model}

\subsection{The transverse-field Ising Hamiltonian}

The model is defined on a chain of $L$ spin-1/2 degrees of freedom by
\begin{equation}
H = -J \sum_i \sigma^z_i \sigma^z_{i+1} - h \sum_i \sigma^x_i ,
\end{equation}
with open boundary conditions throughout this work. We fix $J=1$ and use $h$ as the tuning parameter. For $h<1$ the chain orders ferromagnetically along $z$ (up to the usual two-fold ground-state degeneracy in the thermodynamic limit); for $h>1$ the transverse field dominates and the chain is paramagnetically disordered. A continuous quantum phase transition separates the two phases at $h_c = J = 1$.

\subsection{Entanglement entropy as a diagnostic}

For a bipartition of the chain into a block of $l$ contiguous sites and its complement, the bipartite entanglement entropy $S(l) = -\mathrm{Tr}[\rho_l \ln \rho_l]$ of the ground state is a standard and numerically convenient diagnostic of criticality \cite{Vidal2003}. Away from $h_c$ it saturates to an $l$-independent constant as $l$ grows (the area law for a gapped 1D system); at $h_c$ it grows logarithmically with $l$. For a finite system of length $L$ with open boundaries, conformal field theory gives the Calabrese-Cardy prediction \cite{Calabrese2004}
\begin{equation}
S(l,L) = \frac{c}{6}\ln\!\left[\frac{2L}{\pi}\sin\!\left(\frac{\pi l}{L}\right)\right] + \text{const.},
\label{eq:cc}
\end{equation}
valid asymptotically for $1 \ll l \ll L$, where $c$ is the central charge of the underlying CFT. This formula is the basis for all of the central-charge extractions in Section~\ref{sec:cc}.

\subsection{Quantum-classical correspondence}

The appearance of a 2D CFT in the description of a 1D quantum critical point is not a coincidence particular to this model. Two independent, standard results establish it: (i) the TFIM is exactly related to the two-dimensional classical Ising model via an imaginary-time (Suzuki-Trotter) mapping, with the transverse field $h$ controlling one of the two couplings of the resulting anisotropic classical lattice \cite{Fradkin1978}; the critical points of the two models are identified via the Kramers-Wannier self-duality \cite{Kramers1941} of the classical model. (ii) Independently, a Jordan-Wigner transformation maps the TFIM onto free fermions, whose continuum critical theory is the same free massless Majorana fermion CFT (central charge $c=1/2$) reached by Onsager's classical solution of the 2D Ising model.

\subsection{Scaling near the transition}

Because the TFIM's quantum critical point has dynamical exponent $z=1$, the classical Widom-Kadanoff scaling hypothesis and the 2D Ising critical exponents transfer directly to the quantum problem \cite{Sachdev2011}. The finite-size Calabrese-Cardy formula used throughout Section~\ref{sec:results} --- in which $S(l,L)$ depends on $l$ and $L$ only through the ratio $l/L$ --- is itself an instance of the same generalized-homogeneity idea, applied to entanglement entropy rather than free energy.

\section{Numerical Methods}
\label{sec:methods}

Before settling on the classical pipeline described below, an early stage of this project attempted exact diagonalization directly on the IBM Quantum Platform, for $L=2,3$, and $4$ spins, using both real superconducting-qubit hardware and the accompanying Qiskit simulator backend. This was abandoned in favor of direct classical diagonalization --- not because 2--4 qubits pushed against any hardware qubit-count limit, but because of practical queue and runtime constraints on shared cloud quantum hardware. The simulator backend, in turn, ran into a different and more fundamental limit: being itself a state-vector emulation, it faces exactly the same exponential Hilbert-space wall as any classical exact-diagonalization code, just implemented on different infrastructure. We mention this briefly because it is a useful reminder, in a study organized around the theme of `different numerical methods hit different walls', that the wall encountered can be operational (hardware access and queue time) as much as it is algorithmic (exponential scaling) --- two logically distinct obstacles that are easy to conflate when they are met within the same project.

\subsection{Exact diagonalization (small $L$)}

For $L=8,14$, and $20$, the Hamiltonian was constructed and diagonalized in the full $2^L$-dimensional Hilbert space, giving the exact ground state with no truncation of any kind. The bipartite entanglement entropy at any cut is then obtained directly from the eigenvalues of the reduced density matrix of the exact ground state. Beyond $L\approx 20$--24, the exponential growth of the Hilbert space ($2^{20}\approx 10^6$, doubling with every additional site) makes this approach computationally prohibitive, motivating the switch to a variational, polynomial-cost method.

\subsection{DMRG ($L=40$ to $100$)}

For $L=40,60,80$, and $100$, ground states were obtained via the density-matrix renormalization group (DMRG) \cite{White1992,Schollwock2011} as implemented in the TeNPy library \cite{Hauschild2018}, using a two-site update with a subspace-expansion mixer, initialized from a fully polarized product state, and run to a maximum of 20 sweeps or an energy convergence threshold of $10^{-10}$.

\subsection{Where DMRG is exact, not approximate}
\label{sec:exact}

An often under-emphasized point is that DMRG is only an approximation when the bond dimension $\chi$ used is smaller than the bond dimension a state actually requires. For an open chain of length $L$, an exact matrix-product representation of an arbitrary state requires a bond dimension of up to $2^{L/2}$ at the middle cut. With $\chi_{\max}=200$ used throughout this work, this bound is 16 at $L=8$ and 128 at $L=14$ --- both below 200 --- so the DMRG results at these two sizes are not merely consistent with exact diagonalization, they are exact by construction, identical in principle to full diagonalization. Only from $L=20$ onward (exact bound 1024) does $\chi=200$ constitute a genuine truncation. This gives the small-$L$ overlap between the two methods a stronger status than a numerical cross-check: it is closer to a proof that the two pipelines compute the same object.

\subsection{Convergence diagnostic}

For every $(L,h)$ pair studied --- seven system sizes on an identical 56-point field grid spanning $h=0.50$ to $1.52$, refined to step $0.01$ through $h=0.86$--$0.94$ --- DMRG was run independently at $\chi_{\max}=100$ and $\chi_{\max}=200$, and the resulting mid-chain entropies compared; a discrepancy exceeding $0.01$ was flagged as non-converged. This uniform diagnostic, applied without exception across all 392 $(L,h)$ points in the dataset, is what allows the results below to be presented without excluding or caveating any individual field value.

\subsection{A worked convergence artifact}
\label{sec:artifact}

An earlier, less systematic version of the $L=100$ dataset (computed at a single, fixed $\chi_{\max}=100$, with each field value run independently and no cross-check) showed a fifteen-fold discontinuity in the mid-chain entropy between $h=0.88$ ($S_{\rm mid}=0.049$) and $h=0.90$ ($S_{\rm mid}=0.752$). Re-running the full sweep with the $\chi=100$ vs.\ $\chi=200$ diagnostic above resolves this: at $\chi=200$, $h=0.88$ gives $S_{\rm mid}=0.742$, consistent with its neighbors and with the $h=0.90$ value, and the two bond dimensions agree to within $\sim 10^{-13}$ at every field value in the corrected sweep. The jump was a bond-dimension artifact --- DMRG converging to an anomalous, under-entangled state in a region of near ground-state degeneracy --- not a physical discontinuity. We include this episode because the failure mode (adequate convergence everywhere except in a narrow, easily-missed window near the transition) is generic to DMRG studies of ordered phases and is, in our experience, more often silently present in published finite-size datasets than discussed.

\section{Results}
\label{sec:results}

\subsection{Finite-size flow of the entropy}

\begin{figure}[htbp]
\centering
\includegraphics[width=0.75\textwidth]{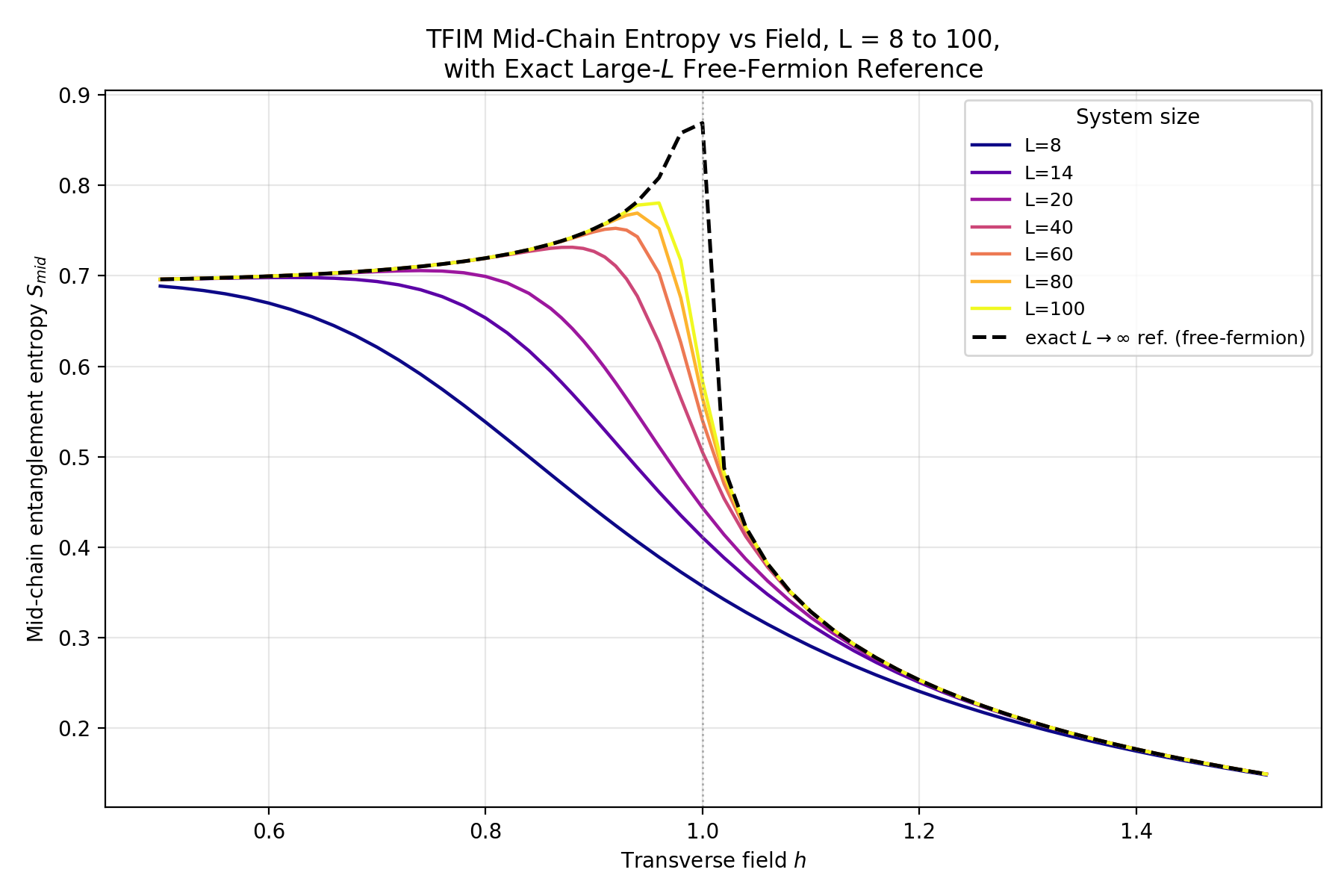}
\caption{Mid-chain entanglement entropy vs.\ transverse field $h$, $L=8$ through 100, $\chi_{\max}=200$ throughout. Dashed black curve: exact large-$L$ reference from the free-fermion solution of the TFIM (see text).}
\label{fig:smid}
\end{figure}

Figure~\ref{fig:smid} shows the mid-chain entanglement entropy against $h$ for all seven system sizes. Away from $h=1$ the curves for different $L$ nearly coincide; approaching $h=1$, larger $L$ develops a taller, sharper, and progressively narrower peak --- the standard finite-size signature of approach to a genuine thermodynamic-limit critical point.

As an independent check on this finite-size flow, we also compute an exact large-$L$ reference curve directly via the free-fermion (Bogoliubov--de Gennes) solution of the TFIM in its Jordan-Wigner-fermionized form, rather than via DMRG \cite{Lieb1961}. This method is polynomial in $L$ and numerically exact, allowing $L$ up to a few thousand; we validated it against independent sparse exact diagonalization (agreement to $\sim 10^{-9}$) and against our own DMRG data at $L=100$ (agreement to $\sim 10^{-11}$) before using it. The resulting dashed curve in Figure~\ref{fig:smid} confirms that the finite-$L$ curves are converging toward the true thermodynamic-limit behavior, not merely toward each other. One subtlety is worth noting: at moderate $L$ (of order 1000), the entropy exactly at $h=1$ can briefly sit below that of its near-critical neighbors, since off-critical entropy saturates quickly with $L$ while critical entropy grows only logarithmically; $h=1$ only becomes the strict global maximum once $L$ is large enough (empirically, $L\gtrsim 3000$ here) for that slow growth to catch up --- a clean illustration of how finite-size effects operate differently exactly at a critical point than nearby.

\subsection{Bond-resolved entropy profile ($L=100$)}

\begin{figure}[htbp]
\centering
\includegraphics[width=0.75\textwidth]{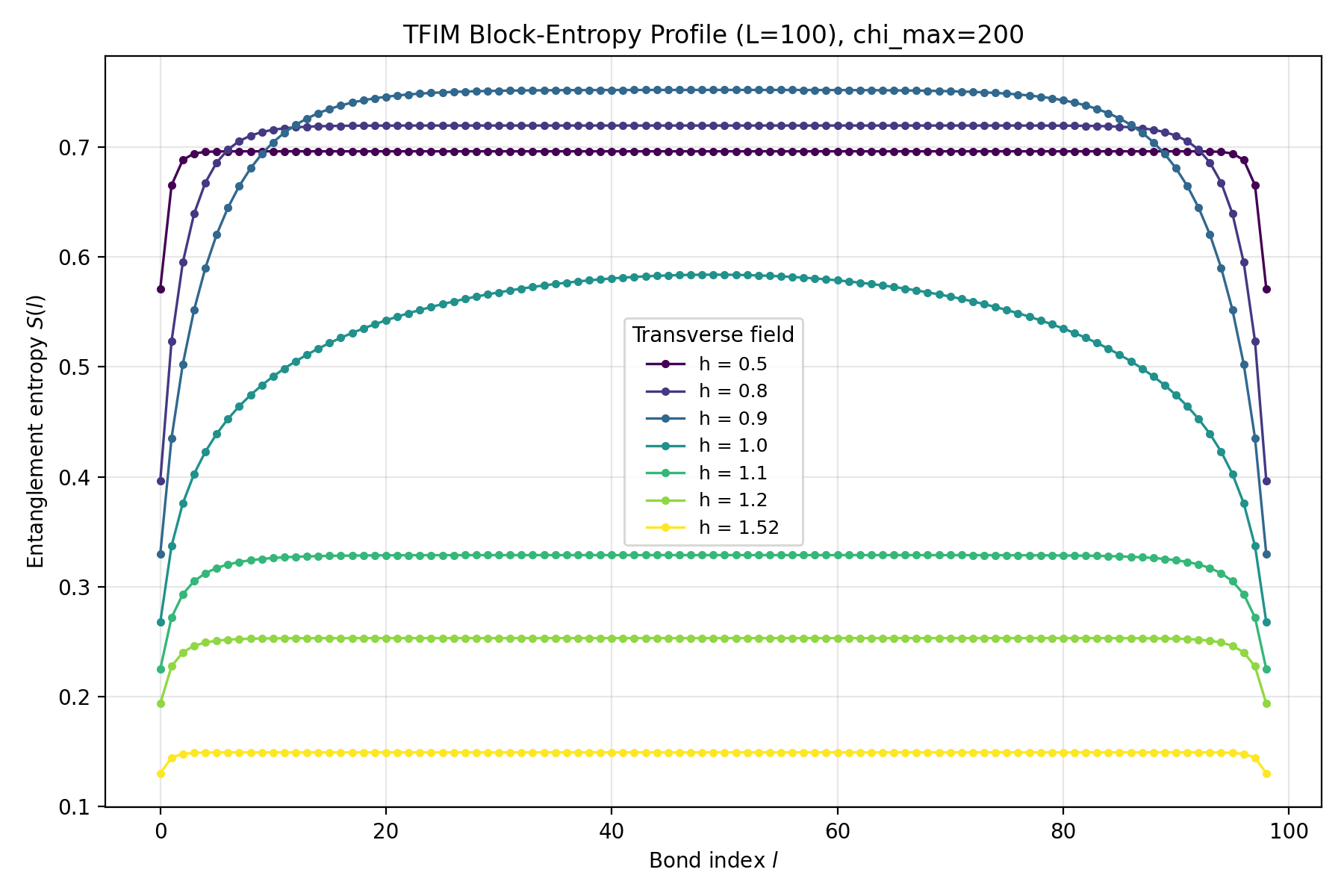}
\caption{Entanglement entropy $S(l)$ vs.\ bond index $l$, $L=100$, for seven representative fields spanning both phases and the transition.}
\label{fig:profile}
\end{figure}

\begin{figure}[htbp]
\centering
\includegraphics[width=0.75\textwidth]{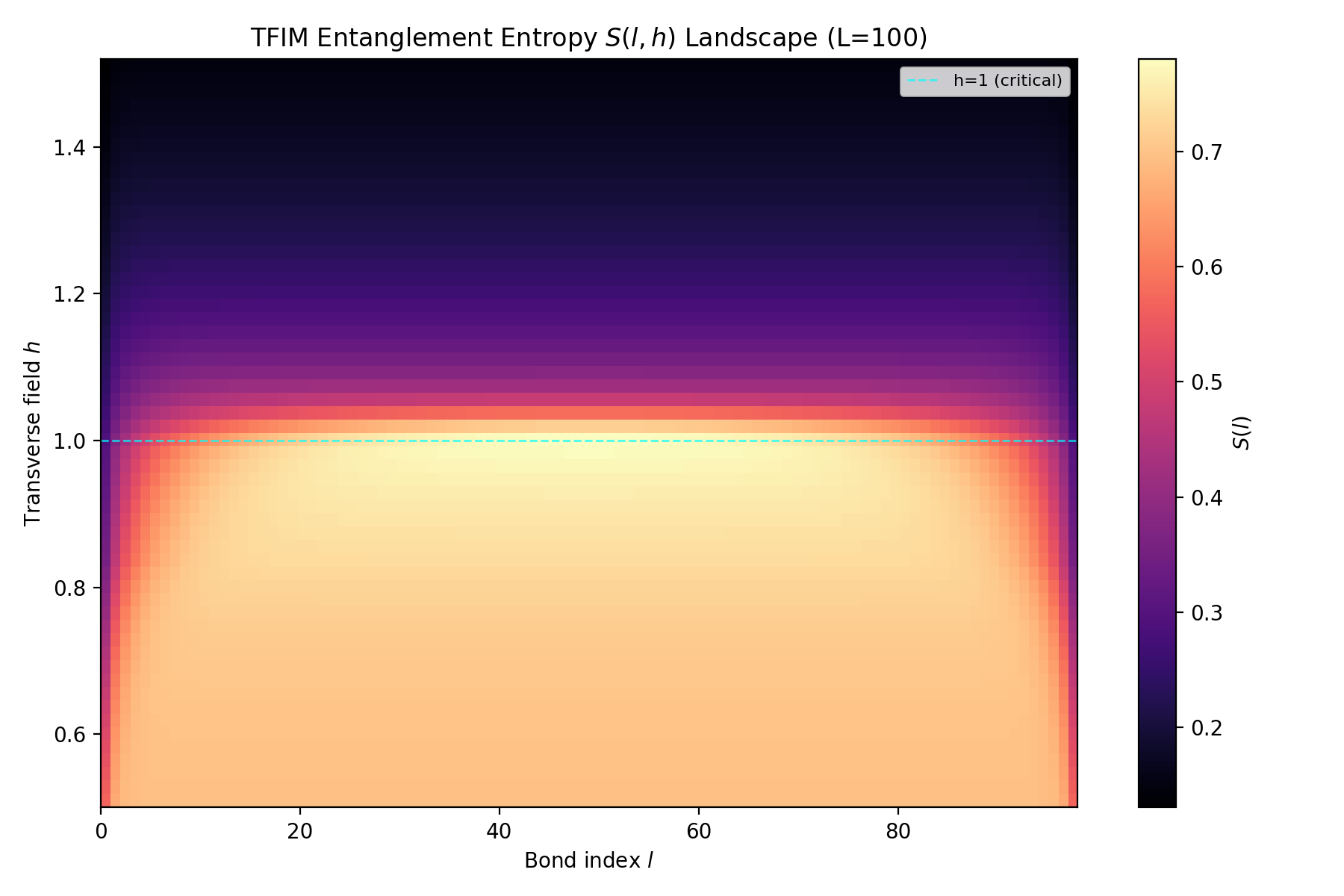}
\caption{Entanglement entropy landscape $S(l,h)$, $L=100$, over the full sampled field range.}
\label{fig:landscape}
\end{figure}

Figures~\ref{fig:profile} and \ref{fig:landscape} show the full bond-by-bond entropy profile at the largest system size studied. Deep in either phase the profile is flat (area law); approaching $h=1$ it develops a pronounced mid-chain arch, the qualitative fingerprint of proximity to criticality in a finite system.

\subsection{Central charge extraction}
\label{sec:cc}

A first estimate uses only the mid-chain entropy from each system size at $h=1$ (Table~\ref{tab:smid}) and the leading logarithmic form $S_{\rm mid}(L) = (c/6)\ln(L) + \text{const.}$; a linear fit across the seven sizes gives slope $0.0893$ ($R^2=0.9996$), i.e.\ $c_{\rm eff}=0.536$.

\begin{table}[htbp]
\centering
\begin{tabular}{cc}
\toprule
$L$ & $S_{\rm mid}(h=1.0)$, $\chi_{\max}=200$ \\
\midrule
8   & 0.357161 \\
14  & 0.411019 \\
20  & 0.443799 \\
40  & 0.505246 \\
60  & 0.540291 \\
80  & 0.564897 \\
100 & 0.583874 \\
\bottomrule
\end{tabular}
\caption{Mid-chain entropy at $h=1.0$ for each system size studied.}
\label{tab:smid}
\end{table}

\begin{figure}[htbp]
\centering
\includegraphics[width=0.7\textwidth]{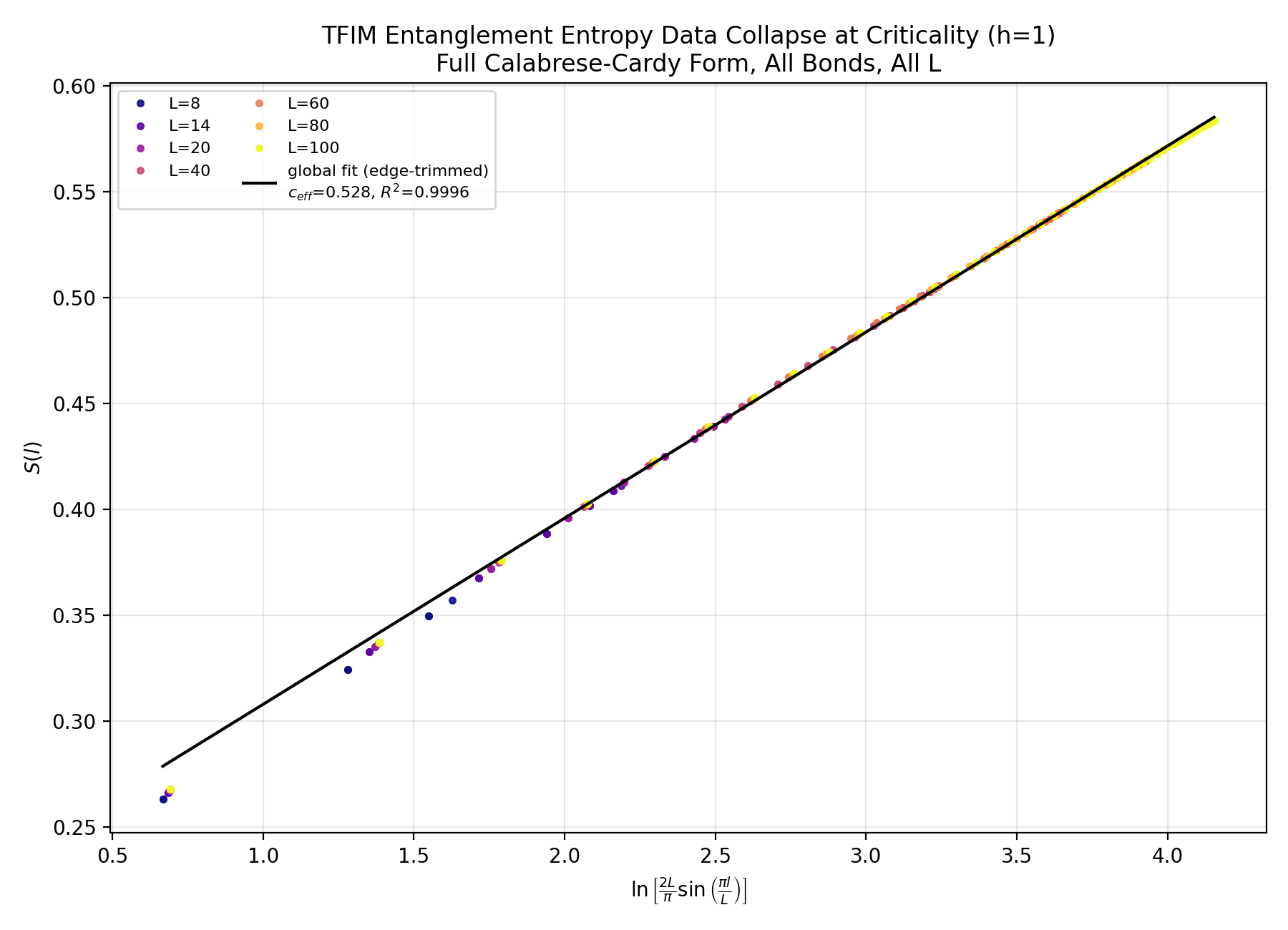}
\caption{Entanglement entropy vs.\ the full Calabrese-Cardy variable, all bonds, all seven system sizes, at $h=1$.}
\label{fig:collapse}
\end{figure}

This uses only seven data points, one per system size. Since the complete bond-by-bond profile was retained at every field value, the same fit can be redone using all 315 available $(l,L)$ pairs at $h=1$ simultaneously, against the full Calabrese-Cardy variable $x = \ln[(2L/\pi)\sin(\pi l/L)]$. The pooled fit gives $c_{\rm eff}=0.541$ ($R^2=0.9991$); more strikingly, all seven system sizes collapse onto a single line when plotted against $x$ (Figure~\ref{fig:collapse}), a direct visual confirmation of conformal finite-size scaling across the entire size range studied.

\begin{figure}[htbp]
\centering
\includegraphics[width=0.7\textwidth]{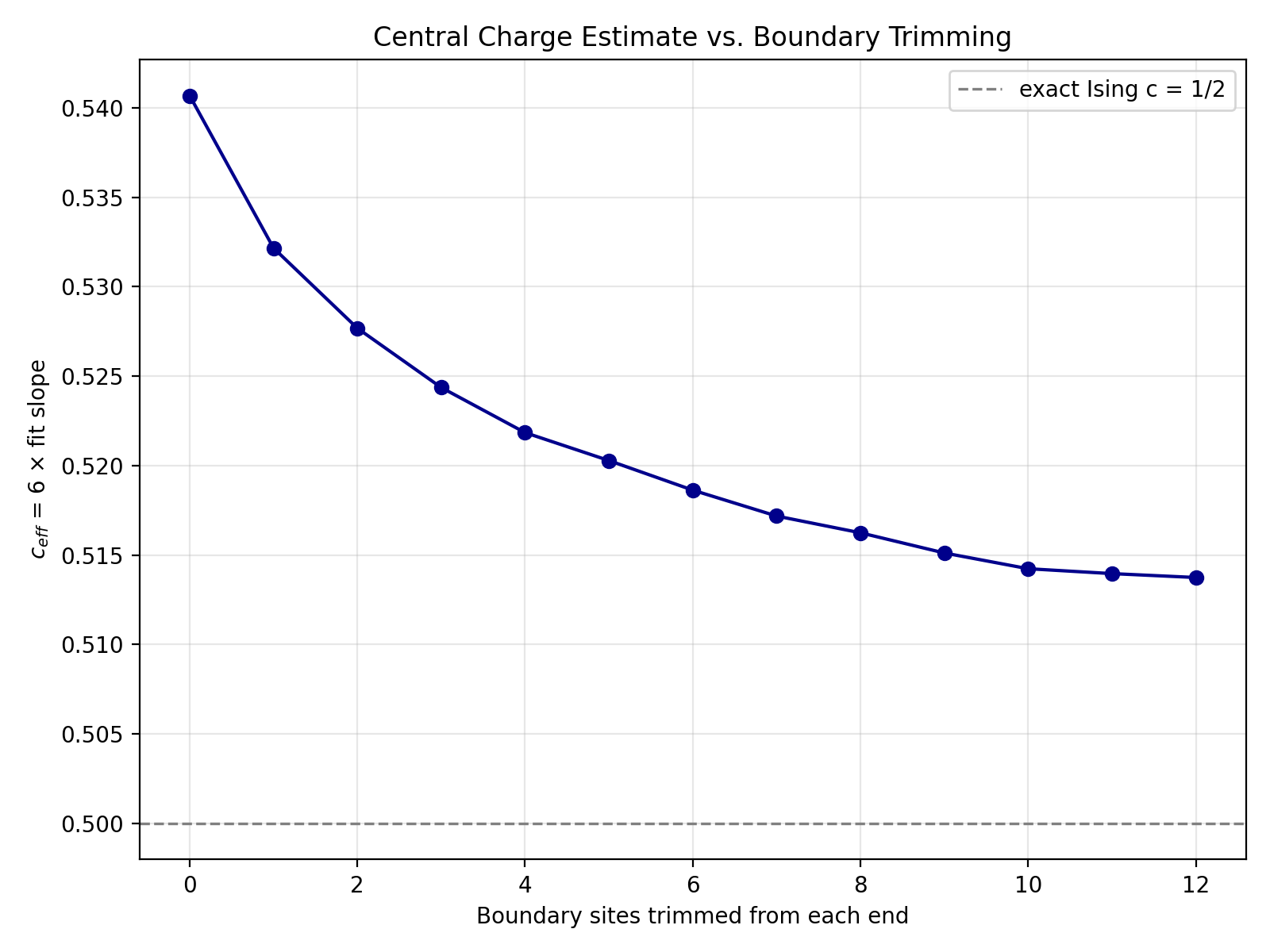}
\caption{Effective central charge vs.\ number of boundary sites excluded from the fit, at $h=1$. Dashed line: exact 2D Ising value $c=1/2$.}
\label{fig:trimming}
\end{figure}

The residual $\sim 8\%$ discrepancy from $c=1/2$ is expected: bonds near the open boundary are subject to non-universal short-distance lattice corrections not captured by the continuum formula. Progressively excluding bonds within a growing distance of either boundary and re-fitting the remainder (Figure~\ref{fig:trimming}) removes this contamination directly: $c_{\rm eff}$ falls monotonically from $0.541$ (no trimming) to $0.514$ once the outermost ten sites on each end are excluded, while the fit quality improves in step ($R^2$ from $0.9991$ to $0.99997$).

\section{Discussion}
\label{sec:discussion}

The dataset presented here is, to our knowledge, unusual chiefly in its internal consistency rather than in any single result: all seven system sizes, from $L=8$ (exact) to $L=100$ (DMRG), were produced with one method, one field grid, and one convergence diagnostic, with zero flagged non-convergences across 392 field values. The central-charge trend of Figure~\ref{fig:trimming} --- monotonic convergence toward $c=1/2$ as an identifiable, physically-motivated source of bias (short-distance boundary lattice corrections) is removed --- is, in our view, more convincing evidence for 2D Ising conformal symmetry in this dataset than any single quoted $c_{\rm eff}$ value could be on its own.

We note one subtlety encountered but not pursued in depth here. Attempting an independent central-charge estimate from the field-dependence of the saturated (gapped-phase) entropy rather than its size-dependence, we found the ferromagnetic and paramagnetic branches gave inconsistent slopes ($c_{\rm eff}\approx 0.2$ and $\approx 0.65$--$0.7$ respectively) under a simple single-log fit, even after replacing the naive critical-scaling correlation length with the exact TFIM result of Pfeuty \cite{Pfeuty1970}. We attribute this to the open-chain ferromagnetic ground state being a symmetric (``cat-state'') superposition of the two nearly-degenerate ordered configurations, which is known to carry additional entanglement beyond the single-vacuum asymptotic formula used here \cite{Calabrese2010}. We leave a careful treatment of this asymmetry, and the corresponding two-branch fit with an explicit cat-state correction, to future work.

\section{Conclusion and Outlook}
\label{sec:conclusion}

We have presented a complete, methodologically consistent numerical study of the transverse-field Ising chain spanning exact diagonalization and DMRG, with an explicit account of where the two methods must agree exactly, a worked example of diagnosing and resolving a DMRG convergence artifact, and a central-charge extraction strengthened by using the complete bond-resolved entropy data rather than a single-cut scan. The results are consistent with the expected 2D Ising universality class throughout: critical field $h_c=1$, and $c_{\rm eff}$ converging toward $1/2$ as boundary lattice corrections are excluded.

Natural extensions include:
\begin{itemize}
\item A systematic finite-size-correction fit (e.g.\ $c_{\rm eff}(\text{edge}) = c + A/\text{edge}$, extrapolated to infinite trimming) to quote $c$ with a proper error bar rather than reading a value off the trimming trend.
\item A careful two-branch (ferromagnetic/paramagnetic) treatment of the gapped-phase entropy saturation, accounting explicitly for the cat-state contribution on the ordered side.
\item Extension of the same consistent ED-to-DMRG pipeline to related models (e.g.\ the XY chain, or the TFIM with an added longitudinal field term, which would provide the direct quantum analog of the classical ordering field).
\end{itemize}

\section*{Acknowledgments}

This work was developed as part of an ongoing quantum sensing and computation research program, with the iSpace industry-university partnership and its institutional collaborators providing broader research context.


\begin{thebibliography}{99}

\bibitem{Pfeuty1970} P.~Pfeuty, ``The one-dimensional Ising model with a transverse field,'' Annals of Physics \textbf{57}, 79--90 (1970).

\bibitem{Lieb1961} E.~Lieb, T.~Schultz, and D.~Mattis, ``Two soluble models of an antiferromagnetic chain,'' Annals of Physics \textbf{16}, 407--466 (1961).

\bibitem{Fradkin1978} E.~Fradkin and L.~Susskind, ``Order and disorder in gauge systems and magnets,'' Phys.\ Rev.\ D \textbf{17}, 2637 (1978).

\bibitem{Kramers1941} H.~A.~Kramers and G.~H.~Wannier, ``Statistics of the Two-Dimensional Ferromagnet,'' Phys.\ Rev.\ \textbf{60}, 252 (1941).

\bibitem{Sachdev2011} S.~Sachdev, \textit{Quantum Phase Transitions}, 2nd ed.\ (Cambridge University Press, 2011).

\bibitem{Vidal2003} G.~Vidal, J.~I.~Latorre, E.~Rico, and A.~Kitaev, ``Entanglement in Quantum Critical Phenomena,'' Phys.\ Rev.\ Lett.\ \textbf{90}, 227902 (2003).

\bibitem{Calabrese2004} P.~Calabrese and J.~Cardy, ``Entanglement Entropy and Quantum Field Theory,'' J.\ Stat.\ Mech.\ P06002 (2004).

\bibitem{Calabrese2010} P.~Calabrese, M.~Campostrini, F.~Essler, and M.~Fagotti, ``Parity effects in the scaling of block entanglement in gapped spin chains,'' Phys.\ Rev.\ Lett.\ \textbf{104}, 095701 (2010).

\bibitem{White1992} S.~R.~White, ``Density matrix formulation for quantum renormalization groups,'' Phys.\ Rev.\ Lett.\ \textbf{69}, 2863 (1992).

\bibitem{Schollwock2011} U.~Schollw\"ock, ``The density-matrix renormalization group in the age of matrix product states,'' Annals of Physics \textbf{326}, 96--192 (2011).

\bibitem{Hauschild2018} J.~Hauschild and F.~Pollmann, ``Efficient numerical simulations with Tensor Networks: Tensor Network Python (TeNPy),'' SciPost Phys.\ Lect.\ Notes \textbf{5} (2018).

\end{thebibliography}
\end{document}